\documentstyle[aps]{revtex}
\begin{document}
%\topmargin=0in
%\headheight=0in
%\headsep=0in
%\oddsidemargin=7.2pt
%\evensidemargin=7.2pt
%\footheight=1in
%\marginparwidth=0in
%\marginparsep=0in
%\textheight=235mm
%\textwidth=160mm

%\documentstyle[12pt]{article}
%\setlength{\textwidth}{6in}
%\setlength{\textheight}{8.0in}
%\setlength{\parskip}{0.05in}
%\flushbottom
%\setlength{\baselineskip}{15.2pt}
%\setlength{\baselineskip}{24.2pt}

%
% section.equation numbering style --- off
%\renewcommand{\theequation}{\thesection.\arabic{equation}}

\newcommand\ie {{\it i.e. }}
\newcommand\eg {{\it e.g. }}
\newcommand\etc{{\it etc. }}
\newcommand\cf {{\it cf.  }}
\newcommand\etal {{\it et al. }}
\newcommand{\be}{\begin{eqnarray}}
\newcommand{\ee}{\end{eqnarray}}
\newcommand\Jpsi{{J/\psi}}
\newcommand\M{M_{Q \overline Q}}
\newcommand\mpmm{{\mu^+ \mu^-}}
\newcommand{\jp}{$ J/ \psi $}
\newcommand{\pp}{$ \psi^{ \prime} $}
\newcommand{\ppp}{$ \psi^{ \prime \prime } $}
\newcommand{\dd}[2]{$ #1 \overline #2 $}
\newcommand\noi {\noindent}
\draft
\preprint{LBL-36755}
%\preprint{January 1995}
\title
{QCD Mechanisms for Double Quarkonium and Open Heavy Meson
Hadroproduction$^\star$}
\footnotetext{$^\star$ This work was supported in part by the Director,
Office of Energy Research, Division of Nuclear Physics of the Office of High
Energy and Nuclear Physics of the U. S. Department of Energy under Contract
Number DE-AC03-76SF0098.}

\author{R. Vogt}
\address
{Nuclear Science Division, Lawrence Berkeley Laboratory, Berkeley, CA 94720,
USA \\
and \\
Institute for Nuclear Theory, University of Washington, Seattle, WA 98195,
USA}

\maketitle

\begin{abstract}
Double $J/\psi$ production on the order of 20-30 pb has been
observed by the NA3
collaboration.  These $\psi \psi$ pairs, measured in $\pi^- N$ interactions at
150 and 280 GeV$/c$ and in $pN$ interactions at 400 GeV$/c$,
carry a large fraction of the projectile momentum, $x_{\psi \psi} \geq
0.6$ for the 150 GeV$/c$ beam and $\geq 0.4$ at 280 GeV$/c$.
We examine several sources of $\psi \psi$ pair production within QCD, including
${\cal O}(\alpha_s^4)$ $\psi \psi$ production, leading-twist $b \overline b$
production and decay, and the materialization of heavy-quark Fock states
in the projectile.  We
estimate the production cross section and the single and double $J/\psi$
momentum and mass distributions for each, comparing the results with the NA3
data, and predict $\psi \psi$
production in $pN$ interactions at 800 GeV$/c$, accessible to current
fixed-target experiments.  We also discuss the observable
implications of open heavy meson
pair production from the intrinsic heavy quark Fock states.
\end{abstract}
\newpage

\vspace{0.5cm}
\begin{center}
{\bf I.  Introduction}
\end{center}

The production of charmonium pairs in the same reaction is expected to be
exceedingly rare in QCD.  However, the NA3 collaboration has measured
$\psi \psi$ pair production in multi-muon events in
$\pi^-$Pt interactions at 150 and 280 GeV \cite{Badpi} and in 400 GeV $p$Pt
interactions \cite{Badp}.
The integrated $\psi \psi$ production cross section per nucleon, $\sigma_{\psi
\psi}$, is $18 \pm 8$
pb at 150 GeV and $30 \pm 10$ pb at 280 GeV
for pion-induced and $27 \pm 10$ pb proton-induced interactions at 400 GeV.
This cross section is reduced relative to the single $J/\psi$ production cross
section, $\sigma_\psi$, by $\sigma_{\psi \psi}/\sigma_\psi
\approx (3 \pm 1) \times 10^{-4}$ in the pion-induced events.

The fraction of the
projectile momentum,  $x_{\psi \psi} = p_{\rm lab}^{\psi \psi}/p_{\rm beam}$,
carried by the $\psi \psi$ pair is always
very large, $x_{\psi \psi}
\geq 0.6$ at 150
GeV and $x_{\psi \psi} \geq 0.4$ at 280 GeV.
Additionally, single $J/\psi$ mesons in the $\psi \psi$ pairs have $x_\psi >
0.15$.  When the data from both pion beams are combined,
$\langle x_{\psi \psi}
\rangle = 0.66$ and $\langle x_\psi \rangle = 0.33$.  As we shall show,
perturbative $ g g$ and $q \overline q$ fusion
processes produce central $\psi \psi$ pairs with lower
average momentum fractions\footnote{Single $J/\psi$'s produced
at rest in the $\pi^- N$ center
of mass have laboratory momentum fractions of $x_\psi = 0.18$
at 150 GeV and 0.14 at 280 GeV.  Likewise, assuming $M_{\psi \psi} = 7$ GeV
leads to laboratory fractions $x_{\psi \psi} = 0.42$ at 150 GeV and 0.3
at 280 GeV for $\psi \psi$ pairs at rest in the $\pi^- N$ center of mass.}.
The average invariant mass of the
pair, $\langle M_{\psi \psi} \rangle = 7.4$ GeV, is
well above the $\psi \psi$ production threshold, $2m_\psi = 6.2$ GeV.
In fact, all of the observed events have an invariant mass
greater than 6.7 GeV.  Additionally, the
average transverse momentum of the pair is quite small,
$p_{T,\psi \psi} = 0.9 \pm 0.1$ GeV, suggesting
that $\psi \psi$ pair production is highly correlated, occurring in the same
interaction. The $pN$ events have a somewhat
lower invariant mass, $\langle M_{\psi
\psi} \rangle \approx 6.8$ GeV, with several events near threshold.
The  $x_{\psi \psi}$ distribution was
not reported.

In this paper, we shall investigate a number of leading-twist and higher-twist
$\psi \psi$ pair production models.  Several leading-twist
production mechanisms have been discussed previously, including ${\cal
O}(\alpha_s^4)$ $\psi \psi$ production via $gg$ fusion and $q \overline q$
annihilation \cite{ES,HM,Russ}, $B \overline B$
production and decay, $B \overline B \rightarrow \psi \psi X$ \cite{BHK},
and the production of a $2^{++} c \overline c c \overline c$
resonance that decays into $\psi \psi$ pairs \cite{LL}.
We review double $J/\psi$ hadroproduction
by these leading-twist mechanisms, updating the
results with more recent parameters and nonscaling parton distributions.
These models are unable to produce $\psi \psi$ pairs with large $x_{\psi
\psi}$.

We then turn to a higher-twist mechanism that easily produces fast
$Q \overline Q$ pairs: intrinsic heavy quark components in
the projectile wavefunction \cite{intc}.
In a companion paper, Ref.\ \cite{VB2},
we evaluated the probability for two $J/\psi$'s to originate from an intrinsic
$|\overline u d c \overline c c \overline c \rangle$ Fock state.
This model was successful in reproducing the general features of the data.
In this paper, we also consider several other sources of double
$J/\psi$ production by intrinsic heavy quark states.
An intrinsic $b \overline b$ pair, $| \overline u d b \overline b \rangle$,
could
decay into $\psi \psi$ pairs, as in leading-twist $b \overline b$ production.
Additionally, an intrinsic
$|\overline u d c \overline c b \overline b \rangle$ state
could produce $\psi \psi$ pairs from a $B \rightarrow J/\psi X$ decay
combined with the coalescence of the $c \overline c$ pair into a $J/\psi$.
Finally, we calculate $\psi \psi$ pair production through leading-twist
$c \overline c$ production with the projectile in an intrinsic $c
\overline c$ state.  Examples of all the processes considered are shown in
Fig.\ 1.

In addition to estimating the $\psi \psi$ production cross section for all of
the leading-twist and higher-twist mechanisms, we compare our results to the
combined NA3 pion-induced data and the invariant mass distribution from the
400 GeV proton-induced production.  We also discuss
$\psi \psi$ pair production
by 800 GeV proton beams, measurable in ongoing fixed-target
experiments.  To facilitate comparison with the data, we present all the
momentum distributions in the laboratory frame and normalize the pion-induced
calculations to the combined 150 and 280 GeV data and the proton-induced
calculations to the 400 GeV data.  The $\psi \psi$ pair mass and momentum
distributions are normalized to the pair rate while the $J/\psi$ momentum
distributions are normalized to the single rate, twice the pair rate.

\vspace{0.5cm}
\begin{center}
{\bf II.  Leading-Twist Production Mechanisms}
\end{center}

In this section we shall review the leading-twist QCD predictions for double
quarkonium hadroproduction,
${\cal O}(\alpha_s^4)$ $\psi \psi$ production and $b
\overline b$ production and decay.  We also estimate double $J/\psi$ production
assuming the existence of a $2^{++} c \overline c c \overline c$
resonance.  These processes are sensitive to the parton
distributions in the pion and nucleon.  We use two sets of recent leading-order
parton distribution functions in our calculations, GRV LO, available for both
nucleons \cite{GRV} and pions \cite{GRVpi}, and Duke-Owens 1.1 \cite{DO11} for
the nucleon with Owens I \cite{OW1} for the pion.  The results we present are
calculated for a proton target, but at these energies and with these scale
choices, the $\pi^- p$ cross sections can be $\approx 50$\% larger than the
corresponding $\pi^- N$ cross sections since the contribution from
pion valence quark annihilation is important.  There is no significant
difference between the $pp$ and $pN$ production cross sections.

Since the calculations are leading order, a $K$ factor is usually invoked to
account for next-to-leading order corrections.  For example, in $b \overline b$
production, $K = \sigma(\alpha_s^3)/\sigma(\alpha_s^2) \approx 1.2-2.5$,
depending on the choice of $m_b$ and the scale parameter $\mu_b$ \cite{NDE}.
Indeed, for $\mu_b = m_b$, $\sigma(m_b=4.5
\, {\rm GeV})/\sigma(m_b=5 \, {\rm GeV})
\approx 3.7$ at ${\cal O}(\alpha_s^2)$ while for $m_b=4.75$ GeV,
$\sigma(\mu_b = 0.5m_b)/\sigma(\mu_b = 2m_b) \approx 6$ at
${\cal O}(\alpha_s^2)$ and 3 at ${\cal O}(\alpha_s^3)$.  Thus over a reasonable
range of $m_b$ and $\mu_b$, the leading order $b \overline b$ production cross
section can vary by an order of magnitude at the energies studied here.

There is thus considerable uncertainty in the normalization of the
leading-twist $\psi \psi$ production cross section.  In
our analysis we will quote a representative value of $\sigma_{\psi \psi}$.
The most important test of these models, however, is the shape of the momentum
and invariant mass distributions since these are not strongly
affected by the target type or the scale parameters.  We will find that while
some of these mechanisms can account for the magnitude of $\sigma_{\psi \psi}$
within a factor of 2-3, none of them can fully describe the characteristics
of the momentum and invariant mass distributions.

\vspace{0.5cm}
\begin{center}
{\bf 1. Production at ${\cal O}(\alpha_s^4)$}
\end{center}

Leading-twist double $J/\psi$ production has been calculated to
leading order in $\alpha_s$, ${\cal O}(\alpha_s^4)$, in Ref.\
\cite{ES,HM,Russ}.  In these caclculations, the
$J/\psi$'s are treated as nonrelativistic bound states
with the appropriate spin and angular momentum quantum numbers,
as in the color-singlet model of quarkonium production \cite{BaierR}.
Examples of $gg \rightarrow \psi \psi$ production diagrams are shown in Fig.\
1(a).  There are 36 diagrams in all, some vanish due to color conservation
and others can be related by crossing \cite{HM}.  The $q \overline q
\rightarrow \psi \psi$ diagrams \cite{ES,HM} are shown in Fig.\ 1(b).  After
four-momentum conservation is accounted for,
the ${\cal O}(\alpha_s^4)$ $\psi \psi$ production cross section
may be written as
\be \frac{d \sigma}{dp_T^2 dy_Q dy_{\overline Q}} =  \sum_q
(F_q(x_1)  F_{\overline q}(x_2) + F_{\overline q}(x_1) F_q(x_2)) \frac{d
\hat{\sigma}}{d \hat{t}}|_{q \overline q} + F_g(x_1) F_g(x_2) \frac{d
\hat{\sigma}}{d \hat{t}}|_{gg} \, \, , \ee
where $x_1  =  m_{T, \psi}(e^{y_1} + e^{y_2})/\sqrt{s} $,
$x_2  =  m_{T, \psi}(e^{-y_1} + e^{-y_2})/\sqrt{s} $ and
$m_{T, \psi}^2 = m_\psi^2 + p_T^2$.  The parton momentum distribution
functions, $F(x)=xG(x)$, are evolved to $Q^2 = 4m_\psi^2$ so that when the
renormalization and factorization scales are assumed to be equal, $\alpha_s
\approx 0.21$.

The subprocess cross sections for $\psi \psi$ production to order
${\cal O}(\alpha_s^4)$ from $q\overline q$ annihilation and $gg$ fusion
are \cite{HM}
\be
\frac{d \hat{\sigma}}{d \hat{t}}|_{q \overline q} = \frac{2}{3\pi} \left(
\frac{16 \pi \alpha_s}{3} \right)^4 \frac{|\Psi(0)|^2}{m_\psi}
\frac{|\Psi(0)|^2}{m_\psi} \frac{1}{\hat{s}^4} [3 + 4y - 10z^2 - 24z^2y
-24z^4] \, \, ,
\ee
\be  \frac{d \hat{\sigma}}{d \hat{t}}|_{g g} = \frac{3}{4\pi} \left(
\frac{\pi \alpha_s}{3} \right)^4 \frac{|\Psi(0)|^2}{m_\psi}
\frac{|\Psi(0)|^2}{m_\psi} \frac{\hat{s}^4 \sum_0^4 y^n C_n(z)}{(\hat{t} -
m_\psi^2)^4 (\hat{u} - m_\psi^2)^4} \, \, ,
\ee
where $y = (\hat{t} + \hat{u})/2\hat{s}$, $z = (\hat{t} -
\hat{u})/2\hat{s}$ and $\hat{s} = x_a x_b s$, $\hat{t}  = -x_1
m_{T, \psi} \sqrt{s}e^{-y_1} + m_\psi^2$, and $\hat{u}  = -x_2
m_{T, \psi} \sqrt{s}e^{y_1} + m_\psi^2$.  The $J/\psi$
wavefunction at the
origin, $|\Psi(0)|^2$, is fixed by the electromagnetic decay
width, $\Gamma(J/\psi \rightarrow e^+ e^-)$.
We use the most recent determination,
$|\Psi(0)|^2 = 0.0689$ GeV$^3$ \cite{Schuler}.  The polynomials $C_n(z)$, from
Ref.\ \cite{HM}, are
\be
C_0 & = & \frac{2}{3} (3981312z^{12} + 258048z^{10} - 1186048z^8 + 35584z^6
\nonumber \\
    &   & \mbox{} + 98144z^4 - 13568z^2 + 335) \nonumber \\
C_1 & = & \frac{8}{3} (1935360z^{10} - 596224z^8 - 403840z^6 + 168084z^4
\nonumber \\
    &   & \mbox{} - 14288z^2 + 137) \nonumber \\
C_2 & = & \frac{8}{3} (1134336z^8 - 1010944z^6 + 239008z^4 - 11472z^2 - 1)
\nonumber \\
C_3 & = & \frac{64}{3} (-10176z^6 + 688z^4 + 828z^2 + 5) \nonumber \\
C_4 & = & 26 (16z^4 + 8z^2 + 1) \nonumber \, \, .
\ee

The predicted cross sections for
$\pi^- p \rightarrow \psi \psi$ are 7.4 pb at 150 GeV
and 16.5 pb at 280 GeV, $\approx 30$-50\% of
the measured total.  At 400 GeV, the predicted $pp \rightarrow \psi \psi$
production cross section is 7.2 pb, $\approx 30$\% below the data,
increasing to 26 pb at 800 GeV \footnote{In
Ref.\ \cite{HM}, a smaller value of the wavefunction, $|\Psi(0)|^2 = 0.0387$
GeV$^3$, and a larger value of the strong coupling constant, $\alpha_s = 0.32$,
were used, leading to similar conclusions about the magnitude of the
production cross section.}.

The calculations are compared with the $\pi^- N \rightarrow \psi \psi X$
data in Fig.\ 2.  The results are not strongly
dependent on the choice of parton distributions.  The momentum
distributions are relatively broad, with $\langle x_\psi \rangle = 0.26$ and
$\langle x_{\psi \psi} \rangle = 0.5$ at 150 GeV; $\langle x_\psi \rangle =
0.21$ and $\langle x_{\psi \psi} \rangle = 0.4$ at 280 GeV. Note that
$\langle x_{\psi \psi} \rangle$ is significantly less than required by
the $\psi \psi$ data.  In fact, all the measured pairs have a larger
$x_{\psi \psi}$ than these averages.  Also 20\% of the
single $J/\psi$'s in the pairs have $x_\psi > 0.5$,
larger than predicted by leading order QCD $\psi \psi$ production.
The $pp\rightarrow \psi \psi$ calculations,
shown in Fig.\ 3, predict narrower distributions than the $\pi^- p$
calculations: $\langle x_\psi \rangle =
0.15$ and $\langle x_{\psi \psi}
\rangle = 0.3$ at 400 GeV; $\langle x_\psi \rangle = 0.11$ and $\langle
x_{\psi \psi} \rangle = 0.24$ at 800 GeV.  We find
$\langle M_{\psi \psi} \rangle =  7$ GeV for both projectiles.
This is in good agreement
with the 400 GeV proton-induced data where the pair mass is low.
However, the $\pi^- N$ data suggests that production occurs above threshold.
Thus although the absolute cross sections are not significantly
below the data, the mass and longitudinal momentum distributions from this
model fail to describe the data.

\vspace{0.5cm}
\begin{center}
{\bf 2. Leading-Twist $b \overline b$ Production}
\end{center}

We estimate the $\psi \psi$ pair production cross section
from leading-twist $b \overline
b$ production and decay.  At ${\cal O}(\alpha_s^2)$, $b \overline b$ production
proceeds by $gg$ fusion and $q \overline q$ annihilation, as shown in Fig.\
1(c).  The $b \overline b$ production cross section has the same form as
Eq.\ (1) but the subprocess cross sections are \cite{Ellis}
\be
\frac{d \hat{\sigma}}{d \hat{t}}|_{q \overline q} =
\frac{\pi \alpha_s^2}{9\widehat{m}_b^4} \frac{\cosh(y_b - y_{\overline b}) +
m_b^2/\widehat{m}_b^2}{(1 + \cosh(y_b - y_{\overline b}))^3} \, \, ,
\ee
\be  \frac{d \hat{\sigma}}{d \hat{t}}|_{g g} = \frac{\pi \alpha_s^2}{96
\widehat{m}_b^4} \frac{8 \cosh(y_b - y_{\overline b}) - 1}{(1 +
\cosh(y_b - y_{\overline b}))^3}
\left( \cosh(y_b - y_{\overline b}) + \frac{2m_b^2}{\widehat{m}_b^2} -
\frac{2m_b^4}{\widehat{m}_b^4} \right) \, \, .
\ee
The parton distributions are evaluated at $Q^2 = 4m_b^2$ with $m_b = 4.75$
GeV.  The predicted rates of open $b \overline b$ production are
$\sigma_{b \overline b}^{\rm tot}(\pi^- p) =  230$ pb at
150 GeV and  2 nb at 280 GeV.  We also find $\sigma_{b \overline b}^{\rm
tot}(p p) = 250$ pb at 400 GeV and 1.3 nb at 800 GeV.  As previously discussed,
changing $m_b$ and $Q = \mu_b$ can change the predicted
$\sigma_{b \overline b}^{\rm tot}$ by a factor of five to ten.

Flavor excitation has been considered as a source of $b$ quark production in
Ref.\ \cite{BHK}.  These quark-gluon scattering
graphs are considered part of the $K$ factor which increases the cross section
by $K \approx 2$ \cite{Ellis}, as already discussed.  In fact, if the
uncertainties are varied to produce the maximum production rate,
the $\pi^- p$ production cross section could be
increased to $\approx 10$ nb, not the 100 nb predicted in Ref.\ \cite{BHK}. A
later study \cite{HM} ruled out $b \overline b$ production as a significant
source of $\psi \psi$ pair production due to the small production
cross section, further reduced by the $B \overline B \rightarrow
\psi \psi$ decay.  Indeed, including the square of
the inclusive branching ratio, $BR(B \rightarrow J/\psi X) \approx 1.3$\%,
\cite{PDG} the results are a factor of 10 to 100 below
the data\footnote{We assume that all $b$ quarks produce final states that
can decay to $J/\psi$'s to get an upper bound on the $\psi \psi$ production
cross section and study inclusive $B$ decays.  Choosing an exclusive decay
channel, such as $B
\rightarrow J/\psi K$, reduces the branching ratio.  Other assumptions of how
the final-state hadrons are produced would also reduce the $J/\psi$ rate.}.
Our predicted $\pi^- p \rightarrow \psi \psi$ production cross sections are
then 0.028 pb at 150 GeV and 0.24 pb at
280 GeV while the corresponding $pp \rightarrow \psi \psi$ cross sections
are 0.03 pb at 400 GeV and 0.16 pb at 800 GeV.

Measurements of charm hadroproduction show that the longitudinal momentum
distribution of final-state $D$ mesons are similar to the predicted charmed
quark distributions.  Using a fragmentation function that reproduces $D$
production in $e^+ e^-$ annihilation \cite{Pete} softens the $D$ distribution
further.  However, in hadroproduction, comoving light partons can combine
with the charmed quarks without significant momentum loss by the charmed quark
so that the fragmentation can be described by a delta function, $D_{D/c} (z)
\approx \delta(z -1)$, in matter \cite{VBH2}.  Since the $b$ quark
fragmentation function from $e^+ e^-$ annihilation predicts less momentum loss
\cite{Chrin}, the delta function approximation of fragmentation should be even
better for $b$ quarks.
Therefore $x_{B \overline B} = x_{b \overline b}$ and $x_B = x_b$.
However, the mass distributions are kinematically
affected since the $B$ meson mass
reduces the available phase space for high mass $B \overline B$
pairs.  We find that the
average $B \overline B$ mass is approximately 1 GeV larger than the average $b
\overline b$ mass and that both increase with incident energy. This predicts
%$B \overline B $ decays result in
a larger $\psi \psi$ invariant mass than ${\cal O}(\alpha_s^4)$
$\psi \psi$ production
since each $J/\psi$ is produced by an isotropic decay, leading
to a larger rapidity gap between the $J/\psi$'s and thus higher $\psi \psi$
pair masses.  Note that the $\psi \psi$ pair masses are larger if the $b$
quarks are assumed to decay into $J/\psi$'s directly.  However, the single and
pair $J/\psi$ distributions are only weakly dependent on whether the parent
particle was a $b$ quark or a $B$ meson.  The $x_\psi$ and $x_{\psi \psi}$
$J/\psi$ distributions have larger average values and narrower widths than
the ${\cal O}(\alpha_s^4)$ calculations.

We calculate the $B \rightarrow J/\psi X$ decay distributions
by a Monte Carlo using $J/\psi$ momentum distributions calibrated to the
$B$ decays measured at CLEO \cite{CLEO}.  The $J/\psi$
and $\psi \psi$ distributions are not smooth due
to the finite statistics of the Monte Carlo.

The $B$ and $J/\psi$ laboratory frame distributions are given in
Figs.\ 4 and 5.  The $B$ and $B \overline B$ longitudinal momentum
distributions are calculated with both sets of parton distribution
functions.  The
invariant mass, $J/\psi$ and $\psi \psi$ momentum distributions are shown for
the GRV
LO partons only to illustrate the differences arising from assuming either
that the $b$
quarks decay directly to $J/\psi$ or that the quarks first hadronize and then
decay.  The average momentum fractions and invariant masses of the $B$ mesons
and $J/\psi$'s
calculated with the GRV LO parton distributions are shown in Table I.
In addition to the shape differences between the calculations and the data, the
predicted normalization is a factor of $\approx 100$
smaller than required by the data. Thus we can conclude that leading-twist
$b \overline b$ production is not an important source of $\psi \psi$
production at these energies.

\vspace{0.5cm}
\begin{center}
{\bf 3. $2^{++} c \overline c c \overline c$ Resonance Production}
\end{center}

In the resonance model of Ref.\ \cite{LL}, $\psi \psi$ pairs with
$M_{\psi \psi} \equiv 7$ GeV are produced by a Drell-Yan type
mechanism where two gluons fuse in a color vector-meson dominance model.
Each gluon couples to a
$c \overline c$ pair in the color-octet-vector representation.  The $c
\overline c$ pairs then couple to a $2^{++} c \overline c c \overline c$
state which subsequently decays to $\psi \psi$ pairs.
The basic process, which has been generalized
to a number of vector-meson final states, is shown in Fig.\ 1(d).  We follow
Ref.\ \cite{LL} in our calculations.
The $\psi \psi$ pair longitudinal momentum
distribution in the laboratory frame is
\be \frac{d\sigma}{dp_{\psi \psi}} = \frac{1}{sp_L} \int_{W^2_{\rm min}}^s
dW^2 \frac{2}{\sqrt{z^2 + W^2/p_L^2}} G_g(x_1) G_g(x_2) \sigma_{gg} \, \, ,
\ee where $G_g(x)$
represent the gluon number densities, $z$ is the fraction of the laboratory
momentum, $p_L$, carried by the $\psi \psi$ pair, $W$ is the center of mass
energy of the two gluons, and $W^2_{\rm
min} = 4m_\psi^2$.  The gluon fusion cross section is
\be \sigma_{gg} = \frac{1}{64} \frac{k}{128 \pi W} \frac{7}{3} \left( 1 +
\frac{2k^2}{3m_\psi^2} + \frac{2k^4}{15m_\psi^4} \right) \frac{a_{\psi \psi}^2
b_{\psi \psi}^2}{(W-M_{\psi \psi})^2 + \Gamma_{\psi \psi}^2(W)/4} \, \, , \ee
where $a_{\psi \psi} = a/\sqrt{3}$ is the resonance decay constant,
$b_{\psi \psi} = \sqrt{2/3} \alpha_s a (4\pi/
f_{\underline \psi})^2$ is the coupling between the gluons and the $2^{++}$
state, $a \simeq \sqrt{30}$, and $(4\pi/f_{\underline
\psi})^2 \simeq 0.02$, the color vector dominance constant.  The width
of the $\psi \psi$ resonance is \be
\Gamma_{\psi \psi}(W) = \frac{a^2 k}{24 \pi} \left( 1 +
\frac{2k^2}{3m_\psi^2} +
\frac{2k^4}{15m_\psi^4} \right) \, \, , \ee where $k = 0.5 \sqrt{W^2 -
4m_\psi^2}$ is the $J/\psi$ momentum in the center of mass.
A single $J/\psi$ in the $\psi \psi$ pair has the laboratory
momentum distribution
\be \frac{d\sigma}{dp_\psi} = \frac{1}{s} \int_{W^2_{\rm min}}^s dW^2
\int dz \frac{2 \sqrt{1-v_F^2}}{\sqrt{z^2 + W^2/p_L^2}} \, G_g(x_1) G_g(x_2) \,
\frac{1}{k} \frac{d\sigma_{gg}}{d\cos \theta} \, \, , \ee
where $v_F = z/\sqrt{z^2 + W^2/p_L^2}$ and the limits on $z$ are determined
from the angle, $\theta$, between the $J/\psi$'s, $-1\leq W(p_\psi
- zp_L/2)/(k \sqrt{z^2p_L^2 + W^2}) \leq 1$.
The angular distribution resulting from the decay of the $2^{++}$ state is
\be \frac{d\sigma_{gg}}{d\cos\theta} & = &
\frac{1}{64} \frac{k}{128 \pi W} \left( \frac{7}{3} +
\frac{k^2}{3m_\psi^2} \left[ \frac{19}{3} - 5 \cos^2 \theta \right]
\right. \\ \nonumber
&   & \mbox{} \left. + \, \frac{k^4}{m_\psi^4} \left[ \frac{5}{9} - \frac{4}{3}
\cos^2\theta +
\cos^4\theta \right] \right) \frac{a_{\psi \psi}^2
b_{\psi \psi}^2}{(W-M_{\psi \psi})^2 + \Gamma_{\psi \psi}^2(W)/4} \, \, . \ee

In the original model of Ref.\ \cite{LL}, scale invariant parton distribution
functions with $\alpha_s = 0.18$ were used.  Repeating the calculation using
the same gluon
distributions, we find agreement with their results, within one standard
deviation of the measured cross sections\footnote{The gluon convolution was
assumed to be $G_{g_1}^{h_1}(x_1) G_{g_2}^{h_2}(x_2) +
G_{g_2}^{h_1}(x_1) G_{g_1}^{h_2}(x_2)$, twice as large as usual
\cite{Ellis}. For consistency with Ref.\ \cite{LL}, we included this
factor in our calculations.  However, this
difference should be noted.}.
However, if the GRV LO gluon distributions are used at $Q=M_{\psi \psi} = 7$
GeV, the
cross sections are reduced significantly.  We find 1.7 pb at 150 GeV and
10 pb at 280 GeV for pion-induced production; 12.8 pb at 400 GeV and 55.2 pb
at 800
GeV from proton-induced production.
The momentum and mass distributions are shown in Figs.\ 6 and 7.
Although the model predicts
narrower momentum distributions than the ${\cal O}(\alpha_s^4)$ process,
the averages are similar.  However,
neither mechanism produces fast $\psi \psi$ pairs.

\vspace{0.5cm}
\begin{center}
{\bf III. Higher-Twist Production: Intrinsic Heavy Quark Mechanisms}
\end{center}

In the following sections, we will discuss
the production of one or two $J/\psi$'s
from intrinsic heavy quark states in the projectile wavefunction.
The wavefunction of a hadron in QCD can be represented
as a superposition of Fock states of quarks and gluons, {\it e.g.}\
the $\pi^-$ wavefunction includes
$\vert \overline  u  d \rangle$, $\vert \overline u d g
\rangle$, $\vert \overline u d Q \overline Q \rangle$, {\it etc.}\ components.
(Since our calculations are applied to pion and proton projectiles, we use
the notation $|n_V Q \overline Q \rangle$ for the heavy quark Fock state
where $n_V \equiv \overline u d$ for a $\pi^-$ and $uud$ for a proton.)
When the projectile scatters in the target, the coherence of the Fock
components is broken and its fluctuations can hadronize \cite{intc,BHMT}.
For example, intrinsic $c \overline c$ fluctuations can be liberated,
provided the system is probed
during the characteristic time, $\Delta t = 2p_{\rm lab}/M^2_{c \overline c}$,
that such fluctuations exist.

Microscopically, the intrinsic heavy quark Fock components
%$|\overline u d Q \overline Q \rangle$,
are generated by
virtual interactions such as $g g \rightarrow Q \overline Q$ where the gluons
couple to two or more projectile valence quarks. The
probability to produce $Q \overline Q$ fluctuations
scales as $\alpha_s^2(m_Q^2)/m_Q^2$ relative to leading-twist production
\cite{BH}.  Thus intrinsic heavy quark production is higher twist.

The dominant Fock configurations are not far off
shell and thus have minimal invariant mass, $M^2 = \sum_i m_{T, i}^2/ x_i$,
where $m_{T, i}^2 =k^2_{T,i}+m^2_i$ is the transverse mass of the
$i^{\rm th}$ particle in the configuration.  Intrinsic $Q \overline Q$
Fock components correspond to configurations with equal rapidity constituents.
Thus, unlike sea quarks generated
from a single parton, intrinsic heavy quarks carry a
large fraction of the parent momentum \cite{intc}. In fact, if the intrinsic
$Q \overline Q$ coalesces into a
quarkonium state, the momentum of the two heavy quarks is combined and this
final state carries a dominant fraction of the projectile momentum.

There is substantial circumstantial evidence for
intrinsic $c \overline c$ states.
For example, the charm structure function of the proton measured by EMC
is significantly larger than predicted
by photon-gluon fusion at large $x_{Bj}$ \cite{EMCic}.
Leading charm production in $\pi N$ and hyperon-$N$ collisions
also requires a charm source beyond
leading twist \cite{769,VB}. The NA3 collaboration has also shown
that single $J/\psi$ production at large $x_F$ is greater
than expected from leading-twist production \cite{Bad}.

The probability distribution
for a general $n$--particle heavy quark Fock state
as a function of $x$ and $\vec{k}_T$ is written
as \be \frac{dP_{\rm iQ}}{\prod_{i=1}^n dx_i d^2k_{T, i}} = N_n \alpha_s^4(M_{Q
\overline Q})
\frac{\delta(\sum_{i=1}^n \vec k_{T, i})
\delta(1-\sum_{i=1}^n x_i)}{(m_h^2 - \sum_{i=1}^n (m_{T, i}^2/x_i)
)^2} \, \, , \ee where
$N_n$, assumed to be slowly varying, normalizes the probability.  The assumed
constituent quark masses are $m_q = 0.3$ GeV, $m_c = 1.5$ GeV, and $m_b = 4.75$
GeV.  While the $\vec k_T$ dependence is
needed to calculate the mass distributions, an average $k_T^2$
can be used to calculate the $x$ distributions.
Thus, eq.\ (11) can be simplified to
\be \frac{dP_{\rm iQ}}{dx_i \cdots dx_n} = N_n \alpha_s^4(M_{Q \overline Q})
\frac{\delta(1-\sum_{i=1}^n x_i)}{(m_h^2 - \sum_{i=1}^n (\widehat{m}_i^2/x_i)
)^2} \, \, , \ee where $\widehat{m}_i^2 =m_i^2 + \langle \vec k_{T, i}^2
\rangle$ is the effective transverse mass.
Assuming $\langle \vec k_{T, i}^2
\rangle$ is proportional to the square of the constituent
quark mass, we adopt the
effective values $\widehat{m}_c = 1.8$ GeV and
$\widehat{m}_q = 0.45$ GeV, as in our
previous work \cite{VBH2,VBH1} and use  $\widehat{m}_b = 5$ GeV.
Note that the resulting $x$ distributions
are independent of the Lorentz frame.  We will show results involving $B$
decays in the laboratory frame.  Eqs.\ (11) and (12) can be
generalized to an arbitrary number of light and heavy partons in the Fock state
so that the probability distribution for the production of {\it e.g.}\
two heavy quark pairs is,
\be \frac{dP_{\rm iQ_1Q_2}}{\prod_{i=1}^n dx_i d^2k_{T, i}} = N_n
\alpha_s^4(M_{Q_1 \overline Q_1}) \alpha_s^4(M_{Q_2 \overline Q_2})
\frac{\delta(\sum_{i=1}^n \vec k_{T, i})
\delta(1-\sum_{i=1}^n x_i)}{(m_h^2 - \sum_{i=1}^n (m_{T, i}^2/x_i)
)^2} \, \, . \ee

The intrinsic $Q \overline Q$ production cross section can be related to the
probability, $P_{\rm iQ}$, by \be \sigma_{\rm iQ}(hN) =
P_{\rm iQ}
\sigma_{h N}^{\rm in} \frac{\mu^2}{4 \widehat{m}_Q^2} \, \, . \ee
In their single $J/\psi$ measurements, the NA3 collaboration separated the
nuclear dependence into a ``hard" contribution with a nearly linear $A$
dependence at low $x_F$ and a contribution scaling as $A^{0.77}$ for $\pi A$
interactions, characteristic of soft interactions, called ``diffractive".
The soft interaction scale parameter, $\mu^2 \sim 0.2$ GeV$^2$, was fixed by
the assumption that the diffractive fraction of the total production cross
section measured by NA3 \cite{Bad}
is the same for charmonium and charmed hadrons.  Therefore, we obtained
$\sigma_{\rm ic}(\pi N) \approx 0.5$ $\mu$b at 200 GeV and $\sigma_{\rm
ic}(p N) \approx 0.7$ $\mu$b \cite{VB}  taking $P_{\rm ic} = 0.3$\% from an
analysis of the EMC charm structure function measurements \cite{EMCic}.

\vspace{0.5cm}
\begin{center}
{\bf 1. Production From $| n_V b \overline b \rangle$ Fock States}
\end{center}

We first examine $\psi \psi$ production from a $| n_V b \overline b \rangle$
configuration.  Because of the scaling properties of the
intrinsic $Q \overline Q$ production probability \cite{VB}
the intrinsic $b \overline b$ production cross section is \be \sigma_{\rm ib}
(hN) = \sigma_{\rm ic}(hN) \left( \frac{\widehat{m}_c}{\widehat{m}_b}
\frac{\alpha_s(M_{b \overline
b})}{\alpha_s(M_{c \overline c})} \right)^4 \, \, . \ee  Thus $\sigma_{\rm ib}
(hN) \approx$ 2-3 nb and
\be \sigma_{\rm ib}^{\psi \psi}(\pi^- N) =
[BR(B \rightarrow J/\psi X)]^2
\sigma_{\rm ib}(\pi^- N) \approx 0.36 \, {\rm pb}
\, \, . \ee  The corresponding proton-induced cross
section, $\sigma_{\rm ib}^{\psi \psi}(pN) \approx 0.51$ pb, is larger.
These results are larger than our quoted leading-twist $b \overline
b$ production cross sections.  However, this difference is strongly dependent
on the leading-twist parameters.  Also, the
intrinsic $b \overline b$ cross section
is proportional to the $hN$ inelastic cross section and changes slowly with
$\sqrt{s}$ while the leading-twist cross section has a steep $\sqrt{s}$
dependence, particularly near threshold.  Although
$b \overline b$ production and decay is frame independent, we compare the
resulting $J/\psi$'s with the data in the laboratory frame.

There are two ways of producing $B$ mesons from the intrinsic $b
\overline b$ pairs.  The first is by standard fragmentation processes.  We will
assume that the momentum of the quark lost through fragmentation is small so
that a delta function can be used as the fragmentation function, as discussed
for leading-twist production.  This assumption works well for
nonleading $D$ meson production \cite{VBH2,VB}.  Then the meson and quark
distributions are identical.
The quark can also coalesce with a projectile valence spectator to
produce mesons if the projectile has the corresponding valence quark.
The coalescence mechanism introduces flavor
correlations between the projectile and the final-state hadrons and produces
$B$'s with a larger fraction of the projectile momentum\footnote{In this
model, with a $\pi^-$
projectile, $B^-(\overline u b)$ and $B^0(d \overline b)$ can be produced by
both coalescence and fragmentation while
$B^+(u \overline b)/\overline B^0(\overline
d b)$ can only be produced by fragmentation.  In a proton projectile, $B^+/B^0$
are produced by coalescence and fragmentation.  We have not made any
assumptions of the relative $B$ meson production rate by coalescence
and fragmentation.}.  Such a coalescence model has been used to successfully
describe $D^-/D^+$ production asymmetries \cite{VB}.
We denote $B$ mesons produced by fragmentation
with $B_F$ and those produced by coalescence with $B_C$.
The production processes are illustrated in Fig.\ 1(e).

If we assume that the $b$ quark fragments into a $B$ meson, the $B$
distribution
is \be \frac{d P_{\rm ib}}{dx_{B_F}} = \int dz
\prod_{i=1}^n dx_i \frac{dP_{\rm ib}}{dx_1 \ldots dx_n}
D_{B/b}(z) \delta(x_{B_F} - z x_b) \, \, , \ee where  $n=4$,5 for pion
and proton projectiles and
$D_{B/b}(z) = \delta(z-1)$ is the $b$ quark fragmentation function.  The
fragmentation mechanism produces $B$ mesons with 25-30\% of the projectile
momentum.  While this is comparable to or larger than the laboratory fractions
carried by $B$'s produced by leading-twist fusion, it is not as large as the
result for $B_C$ production.
The $B_C$ distribution is
\be \frac{d P_{\rm ib}}{dx_{B_C}} = \int \prod_{i=1}^n dx_i
\frac{dP_{\rm ib}}{dx_1 \ldots dx_n}
\delta(x_{B_C} - x_b - x_1) \, \, . \ee  In this case, the $B_C$ can carry
40-50 \% of the projectile momentum.
The increase over $B_F$ production is due to the
light valence quark.  The $B$  and $J/\psi$ distributions are shown in
Fig.\ 8(c) and 8(d) for the pion and Fig.\ 9(c) and 9(d) for the proton.

The $\psi \psi$ pairs can be produced either from
the decays of the $B \overline B$ states
produced by fragmentation of the intrinsic
$b \overline b$ pair, the combination of
fragmentation and coalescence, or, from a
pion projectile only, double coalescence.  The $B_F \overline B_F$
distribution is
\be \frac{d P_{\rm ib}}{dx_{B_F \overline B_F}} = \int dz_b dz_{\overline b}
\prod_{i=1}^n dx_i \frac{dP_{\rm ib}}{dx_1 \ldots dx_n}
D_{B/b}(z_b) D_{\overline B/\overline b}(z_{\overline b})
\delta(x_{B_F} - z_b x_b) \delta(x_{\overline B_F} - z_{\overline b}
x_{\overline b}) \, \, . \ee  These $\psi \psi$ pairs
have the smallest average momentum fractions of all $\psi \psi$ pairs produced
from the
intrinsic $|n_V b \overline b \rangle$
configuration.  If the final-state pair is $B_C \overline B_F$,
\be \frac{d P_{\rm ib}}{dx_{B_C \overline B_F}} = \int dz
\prod_{i=1}^n dx_i \frac{dP_{\rm ib}}{dx_1 \ldots dx_n}
D_{\overline B/\overline b}(z)
\delta(x_{B_C} - x_b - x_1) \delta(x_{\overline B_F} - z x_{\overline b})
\, \, . \ee  An initial $|uud b \overline b
\rangle$ configuration provides no valence antiquark
for $\overline B_C$ production, thus
the $\psi \psi$ pairs resulting from $B_C \overline B_F$ decays carry the
largest fraction of the proton momentum.  However,
when the projectile is a pion, such double coalescence is possible.  All the
pion momentum is transferred to the $B_C \overline B_C$ pair
\be \frac{d P_{\rm ib}}{dx_{B_C \overline B_C}} = \int
\prod_{i=1}^4 dx_i \frac{dP_{\rm ib}}{dx_1 \ldots dx_4}
\delta(x_{B_C} - x_b - x_1) \delta(x_{\overline B_C} - x_{\overline b} - x_2)
\, \, , \ee  {\it i.e.}\ $\langle x_{B_C \overline B_C} \rangle \equiv 1$.
The $B_C \overline B_C \rightarrow \psi \psi X$ decay results in high
momentum $\psi \psi$ pairs.  The $B \overline B$ distributions are shown in
Figs.\ 8(a) and 9(a).

The $B \overline B$ mass distributions can determine the relative probability
for the hadronization mechanism since the available phase space depends on the
number of light quarks involved.  The mass distribution of the
$b \overline b$ quark pair is
\be \frac{dP_{\rm ib}}{dM^2_{b \overline b}} & = &
\int \prod_{i=1}^n dx_i
d^2k_{T, i} \frac{dx_{b \overline b}}{x_{b \overline b}}
d^2k_{T, b \overline b} \frac{dP_{\rm ib}}{\prod_{i=1}^n dx_i d^2k_{T, i}}
\delta(x_{b \overline b} - x_b - x_{\overline b})
\\ \nonumber
&  & \mbox{} \times \delta(\vec k_{T, b} + \vec k_{T, \overline b} - \vec
k_{T, b \overline b}) \,
\delta \left( \frac{M^2_{T, b \overline b}}{x_{b \overline b}} -
\frac{m_{T, b}^2}{x_b} -
\frac{m_{T, \overline b}^2}{x_{\overline b}} \right)
\, \, . \ee   We find $\langle M_{b \overline b} \rangle = 11.75$ GeV.
The corresponding $M_{\psi \psi}$ distributions
are broad with $\langle M_{\psi \psi} \rangle = 8.0$ GeV.
A significant part of the $b \overline b$ mass distribution lies below the $B
\overline B$ production threshold, 10.56 GeV.  In order to compute
$B_F \overline B_F$ production through delta function fragmentation,
we restrict $m_{T,b}$ to be larger than $m_B$.  The mass cut
reduces the yield by a factor of $\sim 20$ and
increases the average mass to $\langle M_{B_F \overline B_F} \rangle = 12.5$
GeV for both projectiles.  This is the largest average  $B
\overline B$ mass because although a mass threshold has
been imposed, the light quark transverse
momentum has not been affected.  The average $\psi \psi$ mass is reduced
to 7.65 GeV, independent of the projectile.

When $B_C \overline B_F$ pairs are created, one of the projectile quarks is
included in the heavy meson pair, \be  \lefteqn{\frac{dP_{\rm ib}}
{dM^2_{B_C \overline B_F}}  = \int \prod_{i=1}^n dx_i
d^2k_{T, i} \frac{dx_{B_C}}{x_{B_C}} dm_{B_C}^2 d^2k_{T, B_C}
\frac{dx_{B_C \overline B_F}}{x_{B_C \overline B_F}}
d^2k_{T, B_C \overline B_F} \theta(m_{B_F} - 5.28 {\rm GeV})} \\
\nonumber &   & \mbox{} \times \frac{dP_{\rm ib}}{\prod_{i=1}^n dx_i
d^2k_{T, i}}
\delta \left( \frac{m^2_{T, B_C}}{x_{B_C}} -
\frac{m_{T, b}^2}{x_{b}} - \frac{m_{T, 1}^2}{x_1}
\right) \delta(\vec k_{T, 1} + \vec k_{T, b} - \vec
k_{T, B_C}) \delta(x_{B_C} - x_{b} - x_{1})  \\ \nonumber
&  & \mbox{} \times\delta \left( \frac{M^2_{T, B_C \overline B_F}}{x_{B_C
\overline B_F}} -
\frac{m_{T, B_C}^2}{x_{B_C}} - \frac{m_{T, \overline B_F}^2}{x_{\overline
B_F}} \right)
\delta(\vec k_{T, B_C} + \vec k_{T, \overline B_F} - \vec
k_{T, B_C \overline B_F}) \delta(x_{B_C \overline B_F} - x_{B_C} - x_{B_F})
\, \, , \ee  reducing the invariant mass of the $B \overline B$ to 12.2 GeV for
pion and 12.4 GeV for proton projectiles.  When the projectile is a pion,
the $B_C \overline B_F$ pair momentum has to be balanced by the remaining
valence quark, resulting in a stronger correlation than in a proton with two
left over valence quarks and a smaller pair invariant mass.  The $\psi \psi$
pair mass is also reduced to 7.5 GeV for a pion projectile and 7.6 GeV for the
proton.

The strongest momentum correlations arise from $B_C \overline B_C$ production
since all the pion momentum is transferred to the meson pair, {\it i.e.}\
$x_{B_C \overline B_C} \equiv 1$ and $\vec k_{T, B_C \overline B_C} \equiv 0$.
The mass distribution for these pairs can be simplified to
\be  \lefteqn{\frac{dP_{\rm ib}}
{dM^2_{B_C \overline B_C}}  =  \int  dx_1 dx_2
\frac{dx_{B_C}}{x_{B_C}(1-x_{B_C})} dm_{B_C}^2 dm_{\overline B_C}^2
d^2k_{T, b} d^2k_{T, \overline b} d^2k_{T, \overline B_C}} \\
\nonumber &   & \mbox{} \times
\delta \left( \frac{m^2_{T, B_C}}{x_{B_C}} -
\frac{m_{T, b}^2}{x_{B_C}-x_1} - \frac{m_{T, 1}^2}{x_1}
\right) \delta \left( \frac{m^2_{T, \overline B_C}}{1-x_{B_C}} -
\frac{m_{T, \overline b}^2}{1-x_{B_C}-x_2} - \frac{m_{T, 2}^2}{x_2}
\right) \\ \nonumber
&  & \mbox{} \times  \delta \left( M^2_{T, B_C \overline B_C} -
\frac{m_{T, B_C}^2}{x_{B_C}} - \frac{m_{T, \overline B_C}^2}{1-x_{\overline
B_C}} \right)
\left( m_\pi^2 - \frac{m_{T, 1}^2}{x_1} -  \frac{m_{T, 2}^2}{x_2} -
\frac{m_{T, b}^2}{x_{B_C}-x_1} - \frac{m_{T, \overline b}^2}{1-x_{B_C}-x_2}
\right)^{-2} \, \, . \ee The $x$ integrations are done in the delta functions,
leaving the $k_T$ and $B$ meson mass integrals to balance the pion invariant
mass, producing the steepest pair mass distribution with $\langle M_{B_C
\overline B_C} \rangle = 11.8$ GeV.  The $\psi \psi$ mass is also reduced to
$\langle M_{\psi \psi}
\rangle = 7.05$ GeV.  The strong phase space restrictions decrease the
relative production probability for this mechanism, making double coalescence
rare.  However, the detection of such states would be a dramatic proof of
the existence of the coalescence mechanism.

With these pair mass distributions, the average $\psi \psi$ momentum is
decreased by 5-10\% with respect to the average obtained assuming that the $b$
quarks decay to $J/\psi$.  The resulting average
momentum fractions and invariant masses are given in Table II.

Although the intrinsic $|n_V b \overline b \rangle$ configuration produces
fast $\psi \psi$ pairs, with 45-80\% of the pion momentum,
the $\psi \psi$ production cross section is nearly a factor of 100 too
small to account for the NA3 data, primarily
due to the small branching ratio for $B \rightarrow J/\psi X$ decay.

\vspace{0.5cm}
\begin{center}
{\bf 2. Production From Doubly Intrinsic $Q \overline Q$ Configurations}
\end{center}

It is possible
for configurations with more than one intrinsic heavy quark pair to exist
in the projectile, shown in Fig.\ 1(f).
Even though the probability for such configurations is
reduced relative to those with only one intrinsic $Q \overline Q$,
the $\psi \psi$ production cross section could be larger from $|n_V c \overline
c c \overline c \rangle$ or $|n_V c \overline
c b \overline b \rangle$ configurations than the $|n_V b \overline b \rangle$
because the square of the $B$
decay branching ratio does not enter.  We now discuss $\psi \psi$
production from such configurations.

\vspace{0.5cm}
\begin{center}
{\bf A. $|n_V c \overline
c c \overline c \rangle$ Configurations}
\end{center}

In this section, we follow the arguments of Ref.\ \cite{VB2}.
The $J/\psi$ production cross section from an $|n_V c \overline c \rangle$
configuration is
\be \sigma_{\rm ic}^\psi (hN) = f_{\psi/h} \sigma_{\rm ic} (hN) \, \, , \ee
where $f_{\psi/h}$ is the fraction of the intrinsic $c \overline c$ pairs
with mass below the $D \overline D$ threshold that coalesce into a final-state
$J/\psi$, $c \overline c \rightarrow J/\psi$.
Recently we estimated $f_{\psi/\pi} \approx 0.03$
and $f_{\psi/p} \approx 0.014$ \cite{VB2} with $m_c = 1.5$ GeV.  The predicted
cross sections are
$\sigma_{\rm ic}^\psi (\pi^- N) = 15$ nb and $\sigma_{\rm ic}^\psi (pN)
= 9.8$ nb, in agreement with the NA3 single $J/\psi$ data \cite{Bad}.
Given the intrinsic charm cross section for
single $J/\psi$ production, we use the $\psi \psi$ data to
estimate the probability to produce two $c \overline c$ pairs
in the projectile,
$P_{\rm icc}$, assuming that all the $\psi \psi$ pair events arise from this
mechanism, as in Ref.\ \cite{VB2}.   Then
\be \sigma_{\rm ic}^{\psi \psi} (hN) =
f_{\psi/h}^2 \frac{P_{\rm icc}}{P_{\rm ic}} \sigma_{\rm ic} (hN)
= f_{\psi/h} \frac{P_{\rm icc}}{P_{\rm ic}} \sigma_{\rm ic}^\psi (hN) \, \, .
\ee
The measured value, $\sigma_{\psi \psi} = \sigma_{\rm ic}^{\psi \psi}
(\pi^- N)
\approx 20$ pb \cite{Badpi}, requires $P_{\rm icc} \approx 4.4\% P_{\rm ic}$,
implying that it is relatively easy to produce a second intrinsic $Q \overline
Q$ state if one is already present.
If $P_{\rm icc}$ is independent of the projectile
identity, $\sigma_{\rm ic}^{\psi \psi}(pN) \approx 6.1$ pb, only 20\% of
the measured cross section at 400 GeV \cite{Badp}.  Note that if part
of the $\pi^- N \rightarrow
\psi \psi$ production cross section is assumed to arise
from leading-twist mechanisms, $P_{\rm icc}$ is reduced.

The single $J/\psi$ distribution in this configuration is
\be \frac{dP_{\rm icc}}{dx_{\psi_1}} = \int \prod_{i=1}^n dx_i
dx_{\psi_2} \frac{dP_{\rm icc}}{dx_1 \ldots dx_n} \delta(x_{\psi_1} - x_{c_1}
-x_{\overline c_1}) \delta(x_{\psi_2} - x_{c_2} - x_{\overline c_2})
\, \, . \ee  We find $\langle x_\psi \rangle = 0.36$ for the pion projectile
and $\langle x_\psi \rangle = 0.33$ for the proton.
The number of single $J/\psi$'s is twice the number of $\psi \psi$ pairs.
The single $J/\psi$'s in the $\psi \psi$ pairs have a lower average momentum
fraction than those
from $|n_V c \overline c \rangle$ Fock states, where $\langle x_\psi
\rangle = 0.62$ for a pion and $\langle x_\psi \rangle = 0.51$ for a proton
\cite{Bad,VBH1}.  The $\psi \psi$ pair distribution is
\be \frac{dP_{\rm icc}}{dx_{\psi \psi}} = \int dx_{\psi_1}
\frac{dP_{\rm icc}}{dx_{\psi_1}} \delta(x_{\psi \psi} -
x_{\psi_1} - x_{\psi_2}) \, \, . \ee
Here we
find $\langle x_{\psi \psi} \rangle = 0.72$ for the pion and
$\langle x_{\psi \psi} \rangle = 0.64$ for the proton.
We compare the frame-independent
calculation to the combined 150 and 280 GeV data.
The $\psi \psi$ pair distributions are in Figs.\ 10(a) and 11(a) while the
single $J/\psi$ distributions are shown in Figs.\ 10(b) and 11(b).

The $\psi \psi$ pair mass distribution from this
configuration is
\be \lefteqn{\frac{dP_{\rm icc}}{dM^2_{\psi \psi}} = \int \prod_{i=1}^n dx_i
d^2k_{T, i}
\prod_{j=1}^2 \frac{dx_{\psi_j}}{x_{\psi_j}} dm_{\psi_j}^2 d^2k_{T, \psi_j}
\frac{dx_{\psi \psi}}{x_{\psi \psi}}
d^2k_{T, \psi \psi} \frac{dP_{\rm icc}}{\prod_{i=1}^n dx_i d^2k_{T, i}}} \\
\nonumber &   & \mbox{} \times
\delta \left( \frac{m^2_{T, \psi_j}}{x_{\psi_j}} -
\frac{m_{T, c_j}^2}{x_{c_j}} - \frac{m_{T, \overline c_j}^2}{x_{\overline c_j}}
\right) \delta(\vec k_{T, c_j} + \vec k_{T, \overline c_j} - \vec
k_{T, \psi_j}) \delta(x_{\psi_j} - x_{c_j} - x_{\overline c_j}) \\ \nonumber
&  & \mbox{} \times\delta \left( \frac{M^2_{T, \psi \psi}}{x_{\psi \psi}} -
\frac{m_{T, \psi_1}^2}{x_{\psi_1}} - \frac{m_{T, \psi_2}^2}{x_{\psi_2}} \right)
\delta(\vec k_{T, \psi_1} + \vec k_{T, \psi_2} - \vec
k_{T, \psi \psi}) \delta(x_{\psi \psi} - x_{\psi_1} - x_{\psi_2}) \, \, , \ee
where $2m_c < m_\psi < 2m_D$.
The delta functions insure conservation of momentum for the single $J/\psi$
constituents and the $\psi \psi$ pair.  The results
are compared with the data in Figs.\ 10(c) and 11(c).
We find $\langle M_{\psi
\psi} \rangle \approx 7.7$ GeV for the pion beam and $\langle M_{\psi \psi}
\rangle \approx 7.4$ GeV for the proton beam.  The smaller $\langle M_{\psi
\psi} \rangle$ in $pN$ interactions is compatible with the data
\cite{Badpi,Badp}.

This mechanism provides a simple frame-independent source of fast $\psi \psi$
pairs as well as fast single $J/\psi$'s and reproduces the strongly
correlated behavior
of the $\psi \psi$ pairs.  The general agreement with the intrinsic charm
model is good.

\vspace{0.5cm}
\begin{center}
{\bf B. $|n_V c \overline c b \overline b \rangle$ Configurations}
\end{center}

We now discuss $\psi \psi$ production from $|n_V c \overline c b
\overline b \rangle$ configurations.  A projectile in such a configuration
could produce a $\psi \psi$ pair from a combination of $B \rightarrow J/\psi$
decays with $c
\overline c \rightarrow J/\psi$ coalescence.
The $\psi \psi$ production cross section is
\be \sigma_{\rm icb}^{\psi \psi}(hN) = f_{\psi/h} \frac{P_{\rm icb}}{P_{\rm
ib}} \sigma_{\rm ib}(hN)
BR(B \rightarrow J/\psi X) \, \, . \ee  If we assume $P_{\rm icb}/P_{\rm ib}
\approx P_{\rm icc}/P_{\rm ic}$, then $\sigma_{\rm icb}^{\psi
\psi}(\pi^- N) \approx 0.044$ pb and $\sigma_{\rm icb}^{\psi \psi}(pN)
\approx 0.020$ pb.  These cross sections are equivalent to those from
leading-twist $b \overline b$ production, much less than the
measured cross sections.  The $\pi N$ cross section is larger since
$f_{\psi/p} < f_{\psi/\pi}$.

The $J/\psi$ distribution from this state is
\be
\frac{dP_{\rm icb}}{dx_\psi} = \int \prod_{i=1}^n dx_i
\frac{dP_{\rm icb}}{dx_1 \ldots dx_n} \delta(x_{\psi} - x_c
-x_{\overline c}) \, \,  \ee  since the intrinsic $c\overline c$
coalescence is
unaffected by the $b$ quark hadronization. We find $\langle x_\psi \rangle
= 0.31$ from a pion projectile and $\langle x_\psi \rangle
= 0.28$ from a proton, independent of frame.
These $J/\psi$ distributions are shown in the dot-dashed curves in
Fig.\ 12(a) and 13(a).

If the $B$ meson is produced by fragmentation,
\be \frac{d P_{\rm icb}}{dx_{B_F}} = \int dz \prod_{i=1}^n dx_i
\frac{dP_{\rm icb}}{dx_1 \ldots dx_n} dx_\psi \delta(x_\psi - x_c -
x_{\overline c}) D_{B/b}(z)
\delta(x_{B_F} - z x_b) \, \, . \ee
Alternatively, if the $b$ quark hadronizes by coalescence,
\be \frac{d P_{\rm icb}}{dx_{B_C}} = \int \prod_{i=1}^n dx_i
\frac{dP_{\rm icb}}{dx_1 \ldots dx_n} dx_\psi \delta(x_\psi -
x_c - x_{\overline c})
\delta(x_{B_C} - x_b - x_1) \, \, . \ee  Again, note that the $B$
production is frame independent but the $J/\psi$ distributions are
presented in the laboratory frame.
The $B_F$ and $B_C$ distributions are given in Figs.\ 12(a) and 13(a)
for the pion and
proton while the $J/\psi$ distributions from the decays
appear in Figs.\ 12(b) and 13(b).

The $\psi \psi$ pairs from the $|n_V c \overline c b \overline b \rangle$ Fock
state arise from a
$B \rightarrow J/\psi X$ decay combined with $c \overline c \rightarrow
J/\psi$ coalescence and boosted
to the laboratory frame.  The average $\psi \psi$ momentum is larger when the
parent $B$ was produced by coalescence.  The results are shown in
Figs.\ 12(c) and 13(c).  The average momentum fractions of the $B$ mesons
and the $\psi \psi$ pairs from this configuration are given in Table III.
While the average longitudinal momentum fraction
is relatively large, the pair mass is sharply peaked near threshold with
$\langle M_{\psi \psi}
\rangle \approx 6.5$ GeV, as shown in Figs.\ 12(d) and 13(d).  Additionally,
the cross section is a factor of 500 below the data.
Therefore it is unlikely that this mechanism could
produce the $\psi \psi$ pairs observed by NA3.

\vspace{0.5cm}
\begin{center}
{\bf 3. Leading-Twist $J/\psi$ Production with an Intrinsic Charm
Configuration}
\end{center}

We have also studied the possibility of double $J/\psi$ production by
simultaneous leading-twist fusion and
coalescence of an intrinsic $c \overline c$ pair, $c \overline c \rightarrow
J/\psi$.
In this case, the parton distributions of the projectile pion and proton
are obtained from the $|n c \overline c \rangle$
Fock configurations, as shown in Fig.\ 1(g), where $n$ represents the valence
quarks plus other light partons.
Thus for $q \overline q$ annihilation with a pion
valence quark, only the
lowest intrinsic charm Fock state, $| \overline u d c\overline c \rangle$,
is needed.  For gluon
fusion, an additional gluon is necessary,
$| \overline u d g c \overline c \rangle$.
Annihilation with a sea quark requires at least a six particle Fock state,
$| \overline u d q \overline q c \overline c \rangle$.
With a proton projectile, the lowest-lying configurations have five, six, and
seven partons.
The valence quark probability distribution from an intrinsic $c \overline
c \rightarrow J/\psi$ state is given by
\be
q_{v_\pi}(x_1) = \frac{dP_{\rm ic}}{dx_1} = \int dx_{2p} dx_c dx_{\overline c}
dx_{\psi_2}
\frac{dP_{\rm ic}}{dx_1 \ldots dx_n} \delta(x_{\psi_2} -x_c - x_{\overline c})
\, \, , \ee
where $x_{2p}$ is a light quark in the projectile (the $2p$ subscript is used
to distinguish it
from the target momentum fraction) and $x_1$ is the projectile parton
contributing to the fusion process, either a valence or sea quark or a gluon.
We also need the $\psi_2$ distribution when $x_1$ is fixed by the fusion
process,
\be
q_{v_\pi}(x_1;x_{\psi_2}) = \frac{dP_{\rm ic}}{dx_1 dx_{\psi_2}}|_{x_1} =
\int dx_{2p} dx_c dx_{\overline c}
\frac{dP_{\rm ic}}{dx_1 \ldots dx_n} \delta(x_{\psi_2} -x_c - x_{\overline c})
\, \, . \ee  The sea quark and gluon distributions are similarly calculated.
The target parton distribution functions are
GRV LO \cite{GRV} and Duke-Owens 1.1 \cite{DO11}.
The projectile and target momentum fractions in the leading-twist fusion
process are $x_{1,2} = (\pm x_{\psi_1} + \sqrt{x_{\psi_1}^2 + 4\tau^2})/2$,
where $2m_c/\sqrt{s} < \tau < 2m_D/\sqrt{s}$.  The $c \overline c$ pair is
assumed to neutralize its color and produce a $J/\psi$ by emitting a soft
gluon without significant momentum loss.

The $J/\psi$ produced by parton fusion has the momentum distribution
\be
\frac{d\sigma}{dx_{\psi_1}} & =  & 2 F f_{\psi/h}
\int_{2m_c/\sqrt{s}}^{2m_D/\sqrt{s}}
\frac{\tau d\tau}{x_2 \sqrt{x_{\psi_1}^2 + 4\tau^2}} [q_{v_p}(x_2)
q_{v_\pi}(x_1) \widehat{\sigma}_{q \overline q \rightarrow c \overline c}
+ G_p(x_2) G_\pi(x_1) \widehat{\sigma}_{gg \rightarrow c \overline c} \nonumber
\\ &   & + q_{s_p}(x_2)
q_{s_\pi}(x_1) \widehat{\sigma}_{q \overline q \rightarrow c \overline c}]
\, \, , \ee
written in order of increasing number of partons in the minimal
projectile Fock state.  We define $x q_{v_p}(x) = xu_p(x) +
x \overline d_p(x)$ for annihilation
with the $\pi^- (\overline u d)$ valence quarks and $x q_{s_p}(x) = x u_p(x) +
x \overline u_p(x) + xd_p(x) + x \overline d_p(x) + 2 x s_p(x)$ for
$q \overline q$ annihilation with pion sea quarks where $xu_p =
xu_{v_p} + xu_{s_p}$ and  $xd_p = xd_{v_p} + xd_{s_p}$.
These $J/\psi$ distributions, shown in Figs.\ 14(a) and
15(a) for pion and proton beams, are central.  In the
laboratory frame, $\langle x_\psi \rangle_{\rm pf} \approx 0.26$ from
$\pi^- p$ interactions and $\langle x_\psi \rangle_{\rm pf} \approx 0.18$
from $pp$ interactions, nearly independent of the beam energy.
The $J/\psi$ distribution from the intrinsic charm state is
\be
\frac{d\sigma}{dx_{\psi_2}} & = & 2 F f_{\psi/h} \int_0^1 dx_{\psi_1}
\int_{2m_c/\sqrt{s}}^{2m_D/\sqrt{s}}
\frac{\tau d\tau}{x_2 \sqrt{x_{\psi_1}^2 + 4\tau^2}} [q_{v_p}(x_2)
q_{v_\pi}(x_1;x_{\psi_2}) \widehat{\sigma}_{q \overline q \rightarrow c
\overline c} \nonumber \\ &   &
+ G_p(x_2) G_\pi(x_1;x_{\psi_2}) \widehat{\sigma}_{gg
\rightarrow c \overline c}
+ q_{s_p}(x_2) q_{s_\pi}(x_1;x_{\psi_2})
\widehat{\sigma}_{q \overline q \rightarrow c \overline c}]
\, \, . \ee
These distributions, shown in the laboratory frame in Figs.\ 14(b) and
15(b), have larger average momentum fractions, $\langle x_\psi \rangle_{\rm
ic} \approx 0.56$ and  0.46 for pion and proton projectiles.  Thus one fast
and one slow $J/\psi$ are produced, in agreement with the single $J/\psi$
data.  The $\psi \psi$ distributions are strongly forward peaked, with
$\langle x_{\psi \psi} \rangle \approx 0.84$ for a pion and, for a proton,
$\langle x_{\psi \psi} \rangle \approx 0.68$, larger than the data.   The
average $\psi \psi$ pair mass is $\approx 8$ GeV.
However, the mass distributions do not appear strongly correlated.

The parameter $F$ is introduced since the leading-twist
fusion cross section includes
all $c \overline c$ resonances below the $D \overline D$ threshold.  When $m_c
= 1.5$, $F \approx 1/5$.
We also include the factor $f_{\psi/h}$ for $J/\psi$ production from
the intrinsic charm state.  These two suppression factors predict a
$\psi \psi$ production cross section of
$\approx 3.5$ pb for $\pi p$ and $\approx 2.5$ pb for $pp$
production.

\vspace{1.0cm}
\begin{center}
{\bf IV. Summary}
\end{center}

A summary of all the predicted cross sections is given in Table IV.
{}From the magnitude of the calculated cross sections alone, it appears that
approximately 50\% of the measured $\psi \psi$
cross section can be
attributed to the leading-twist mechanisms.
However, only the higher-twist
mechanisms produce fast $\psi \psi$ pairs.  The intrinsic heavy
quark states offer a promising alternative, especially since all $\psi \psi$
pairs produced through these mechanisms carry a significant fraction of the
projectile momentum.  Particularly, the $|n_V c
\overline c c \overline c \rangle$ configuration can also account for the
size of the $\psi \psi$ cross section, as discussed in \cite{VB2}.

The correlations in $B \overline B$ production also suggest an intriguing test
of the intrinsic heavy quark mechanism.  In particular, if both $B$'s are
produced by coalescence, all the pion momentum would be transferred to the
$B \overline B$ pair.  The same type of correlation should be observable in
$D \overline D$ production in $\pi A$ interactions.  Measurements of the
$D^-/D^+$ asymmetry could be extended to the momentum and invariant mass
distributions of $D \overline D$ pairs.

We thank S.J. Brodsky, P. Hoyer, G. Ingelman, and W.-K. Tang for stimulating
discussions and J.-C. Peng for encouragement and for providing the $B$ decay
Monte Carlo.

\newpage
\begin{center}
{\bf Figure Captions}
\end{center}
\vspace{0.2in}
\noindent Figure 1.  (a) Examples of ${\cal O}(\alpha_s^4)$ QCD graphs for
$gg \rightarrow \psi \psi$. (b) The diagrams for ${\cal O}(\alpha_s^4)$
$q \overline q \rightarrow \psi \psi$ production.  (c)  Leading-twist $b
\overline b$ production by $q \overline q$ annihilation and $gg$ fusion.
(d) Drell-Yan type resonance production.  The intermediate state
contains four quarks and $\underline \psi$ denotes a  color-octet-vector $c
\overline c$. (e) An intrinsic $b \overline b$ pair illustrating the
coalescence $(B_C)$ and fragmentation $(B_F)$ production processes. (f)  An
example of a pair of intrinsic $Q \overline Q$ states in the projectile.
(g) Diagrams of leading-twist $J/\psi$ production with the projectile in an
intrinsic $c \overline c$ state.  Fusion production by valence quark
annihilation, gluon fusion, and sea quark annihilation are shown respectively.
\vspace{0.2in}

\noindent Figure 2. The ${\cal O}(\alpha_s^4)$ $\pi^- p \rightarrow \psi
\psi$ distributions in the
laboratory frame compared to the NA3 data. The $x_{\psi \psi}$
distributions are shown in (a), the $x_\psi$ distributions in
(b), and the pair mass distribution in (c).  The curves show:
at 150 GeV, GRV LO (solid) and DO 1.1 (dashed); at 280 GeV,
GRV LO (dot-dashed) and DO 1.1 (dotted).
\vspace{0.2in}

\noindent Figure 3.  The ${\cal O}(\alpha_s^4)$ $p p \rightarrow \psi
\psi$ distributions in the
laboratory frame. The $x_{\psi \psi}$
distributions are shown in (a), the $x_\psi$ distributions in
(b), and the pair mass distribution is compared to the NA3 data
in (c).  The curves show: at 400 GeV, GRV LO (solid) and DO 1.1 (dashed);
at 800 GeV, GRV LO (dot-dashed) and DO 1.1 (dotted).
\vspace{0.2in}

\noindent Figure 4.  Leading-twist $\pi^- p \rightarrow b \overline b$
production in the laboratory frame, compared to the NA3 $\psi \psi$ data.
The $x_{b \overline b}$
distributions are shown in (a) and the $x_b$ distributions in (c) with
GRV LO (solid) and DO 1.1 (dashed) at 150 GeV and
GRV LO (dot-dashed) and DO 1.1 (dotted) at 280 GeV.
The remaining results, with the GRV set, compares the assumptions of
$b \overline b$ decay vs.\ $B \overline B$ production and decay.  The solid and
dot-dashed  curves give the $b \overline b$ distributions at 150 and 280 GeV
while the dashed and dotted curves show the $B \overline B$ distributions at
the same energies.  The $x_{\psi \psi}$ and $x_\psi$
distributions are shown in (b) and (d), the
$\psi \psi$ pair mass distribution in (f), and the invariant mass of the
$b \overline b$ and $B \overline B$ pairs in (e).
\vspace{0.2in}

\noindent Figure 5. Leading-twist $pp \rightarrow b \overline b$ production
in laboratory frame.  The $x_{b \overline b}$
distributions are shown in (a), the $x_b$ distributions in (c) with
GRV LO (solid) and DO 1.1 (dashed) at 400 GeV and
GRV LO (dot-dashed) and DO 1.1 (dotted) at 800 GeV.
The remaining results, with the GRV set, compares the assumptions of
$b \overline b$ decay vs.\ $B \overline B$ production and decay.  The solid and
dot-dashed  curves give the $b \overline b$ distributions at 400 and 800 GeV
while the dashed and dotted curves show the $B \overline B$ distributions at
the same energies.  The $x_{\psi \psi}$ and $x_\psi$
distributions are shown in (b) and (d), the
$\psi \psi$ pair mass distribution in (f), compared to the NA3 data,
and the invariant mass of the $b \overline b$ and $B \overline B$ pairs in (e).
\vspace{0.2in}

\noindent Figure 6. The $\pi^- N \rightarrow 2^{++} c \overline c c
\overline c \rightarrow \psi \psi$ distributions
in the laboratory frame, compared to the NA3 data. The $x_{\psi \psi}$
distributions are shown in (a), the $x_\psi$ distributions in
(b), and the $M_{\psi \psi}$ distribution in (c).  The
calculations are: at 150 GeV, GRV LO (solid) and scaling (dashed) gluon
distributions; at 280 GeV, GRV LO (dot-dashed) and scaling (dotted) gluons.
\vspace{0.2in}

\noindent Figure 7.  The $p N \rightarrow 2^{++} c \overline c c
\overline c \rightarrow \psi \psi$ distributions
in the laboratory frame. The $x_{\psi \psi}$
distributions are shown in (a), the $x_\psi$ distributions in
(b), and the $M_{\psi \psi}$ distribution is compared with the NA3 data
in (c).  The
calculations are: at 400 GeV, GRV LO (solid) and scaling (dashed) gluon
distributions; at 800 GeV, GRV LO (dot-dashed) and scaling (dotted) gluons.
\vspace{0.2in}

\noindent Figure 8.  Frame-independent
$b \overline b$ production is calculated from an
intrinsic $| \overline u d b \overline b \rangle$ configuration.
In (a), the $x_{B_F \overline B_C}$
(solid) and $x_{B_F \overline B_F}$ (dashed) distributions
are shown.  If both $B$'s are produced
by coalescence, $x_{B_C \overline B_C} \equiv 1$. In (c), the $x_{B_F}$
(solid) and $x_{B_C}$ (dashed) distributions are shown.
The invariant masses of the $b \overline b$ (solid), $B_F \overline B_F$
(dashed), $B_C \overline B_F$ (dot-dashed), and $B_C \overline B_C$ (dotted)
pairs are shown
in (e).  The $J/\psi$ distributions are in the laboratory frame.  The
$x_{\psi \psi}$ distributions are shown in (b).
The solid and dot-dashed curves show $B_F \overline B_C \rightarrow \psi
\psi$ pairs at 150 and 280 GeV while the dashed and dotted curves show
$B_F \overline B_F \rightarrow \psi \psi$ pairs
at 150 and 280 GeV.  The thin solid and
dashed curves are results from $B_C \overline B_C \rightarrow \psi \psi$
decays at 150 and 280 GeV. The $x_\psi$
distributions are in (d).  The $B_C
\rightarrow J/\psi X$
decays are given in the dashed (150 GeV) and dot-dashed (280 GeV) while
$B_F \rightarrow J/\psi X$ decays are shown in the solid (150 GeV) and
dotted (280 GeV) curves.  Finally, the $M_{\psi \psi}$ distributions
from $B_F \overline B_F$ decays (solid)
and $B_C \overline B_C$ decays (dashed) are given in (f).
\vspace{0.2in}

\noindent Figure 9.  Frame-independent
$b \overline b$ production is calculated from an
intrinsic $|u u d b \overline b \rangle$ configuration.
In (a), the $x_{B_F \overline B_C}$
(solid) and $x_{B_F \overline B_F}$ (dashed) distributions
are shown.  In (c), the $x_{B_F}$
(solid) and $x_{B_C}$ (dashed) distributions are shown.
The invariant masses of the $b \overline b$ (solid), $B_F \overline B_F$
(dashed), and $B_C \overline B_F$ (dot-dashed) pairs are shown
in (e).  The $J/\psi$ results are shown in the laboratory frame.  The
$x_{\psi \psi}$ distributions are shown in (b).
The solid and dot-dashed curves show the
$B_F \overline B_C \rightarrow \psi \psi$ pairs
at 400 and 800 GeV while the dashed and dotted curves
show $B_F \overline B_F \rightarrow \psi \psi$ decays
at 400 and 800 GeV.  The $x_\psi$
distributions are in (d).  The $B_C
\rightarrow J/\psi X$
decays are given in the dashed (400 GeV) and dot-dashed (800 GeV) and the
$B_F \rightarrow J/\psi X$ decays are given by the solid (400 GeV) and
dotted (800 GeV) curves.  Finally, the $M_{\psi \psi}$ distribution
is in (f).
\vspace{0.2in}

\noindent Figure 10. Frame-independent $\psi \psi$ production from a
$|\overline u d c \overline c c \overline c \rangle$ configuration.
The $x_{\psi \psi}$
distribution is shown in (a), the $x_\psi$ distribution in
(b), and the mass distribution in (c).
\vspace{0.2in}

\noindent Figure 11. Frame-independent $\psi \psi$ production from a
$|u u d c \overline c c \overline c \rangle$ configuration.
The $x_{\psi \psi}$
distribution is shown in (a), the $x_\psi$ distribution in
(b), and the mass distribution in (c).
\vspace{0.2in}

\noindent Figure 12.  Results for $J/\psi$ and $\psi \psi$ production from an
intrinsic $| \overline u d c \overline c b \overline b \rangle$ configuration.
In (a) we show the frame-independent results for: $c \overline c \rightarrow
J/\psi$ (solid), $B_C$ (dashed)
and $B_F$ (dot dashed) production.  All other results are in the laboratory
frame. In (b), the $B_C$ decays
are given by the solid and dotted curves for 150 and 280 GeV while the
$B_F$ decays are shown in the dashed and dot-dashed
curves for 150 and 280 GeV.  In (c) and (d) the $x_{\psi \psi}$ and $M_{\psi
\psi}$ distributions result from the combination of $c \overline c
\rightarrow J/\psi$ with a decaying $B$: pairing with a $B_C$ results in
the dashed (150 GeV) and dotted (280 GeV) curves while a  $B_F$ pairing
produces the results given in the solid (150 GeV) and dot-dashed (280 GeV)
curves.
\vspace{0.2in}

\noindent Figure 13.  Results for $J/\psi$ and $\psi \psi$ production from an
intrinsic $| u u d c \overline c b \overline b \rangle$ configuration.
In (a) we show the frame-independent results for: $c \overline c \rightarrow
J/\psi$ (solid), $B_C$ (dashed)
and $B_F$ (dot dashed) production.  All other results are in the laboratory
frame. In (b), the $B_C$ decays
are given by the solid and dotted curves for 150 and 280 GeV while the
$B_F$ decays are shown in the dashed and dot-dashed
curves for 150 and 280 GeV.  In (c) and (d) the $x_{\psi \psi}$ and $M_{\psi
\psi}$ distributions result from the combination of $c \overline c
\rightarrow J/\psi$ with a decaying $B$: pairing with a $B_C$ results in
the dashed (150 GeV) and dotted (280 GeV) curves while a  $B_F$ pairing
produces the results given in the solid (150 GeV) and dot-dashed (280 GeV)
curves.
\vspace{0.2in}

\noindent Figure 14.  $\psi \psi$ pairs produced by
simultaneous
$J/\psi$ production by leading-twist fusion with a pion in an
$|\overline u d n c \overline c \rangle$
configuration, in the laboratory
frame.  The
leading-twist and intrinsic $c \overline c$ $J/\psi$ distributions are given
in (a) and (b), and the $x_{\psi \psi}$ and mass distributions are shown
in (c) and (d).  The
calculations are: at 150 GeV, GRV LO (solid) and DO 1.1 (dashed);
at 280 GeV, GRV LO (dot-dashed) and DO 1.1 (dotted).
\vspace{0.2in}

\noindent Figure 15.  $\psi \psi$ pairs produced by simultaneous
$J/\psi$ production by leading-twist fusion with a proton in an
$|u u d n c \overline c \rangle$
configuration, in the laboratory
frame.  The
leading-twist and intrinsic $c \overline c$ $J/\psi$ distributions are given
in (a) and (b), and the $x_{\psi \psi}$ and mass
distributions are shown in (c) and (d).  The
calculations are: at 400 GeV, GRV LO (solid) and DO 1.1 (dashed);
at 800 GeV, GRV LO (dot-dashed) and DO 1.1 (dotted).
\vspace{0.2in}
\newpage

\mediumtext
\begin{table}
\caption{The average laboratory momentum fractions and invariant masses
for leading-twist
$B \overline B$ production and decay in $hp$ interactions
calculated with the GRV LO parton distributions.  Note that the combined
$\pi^- N \rightarrow \psi \psi X$ data give $\langle x_\psi \rangle = 0.33$,
$\langle x_{\psi \psi} \rangle = 0.66$, and $\langle M_{\psi \psi} \rangle
= 7.4$ GeV while the average $\psi \psi$ mass from the $pN \rightarrow \psi
\psi X$ measurements is $\langle M_{\psi \psi} \rangle = 6.8$ GeV.}
\begin{tabular}{ccccccc}
$h$, $p_{\rm beam}$ (GeV) & $\langle x_B \rangle$ & $\langle x_\psi \rangle$ &
$\langle x_{B
\overline B} \rangle$ & $\langle x_{\psi \psi} \rangle$ & $\langle M_{B
\overline B} \rangle$ (GeV) & $\langle M_{\psi \psi} \rangle$ (GeV)\\
\tableline
$\pi^-$, 150 & 0.34 & 0.24 & 0.65 & 0.50 & 11.5 & 7.1 \\
$\pi^-$, 280 & 0.27 & 0.19 & 0.52 & 0.41 & 12.1 & 7.4 \\
$p$, 400     & 0.19 & 0.13 & 0.37 & 0.29 & 11.9 & 7.3 \\
$p$, 800     & 0.15 & 0.10 & 0.27 & 0.22 & 12.3 & 7.5 \\
\end{tabular}
\end{table}

\widetext
\begin{table}
\caption{The average momentum fractions of $B$ and $B \overline B$ production
and decay
from intrinsic $|n_V b \overline b \rangle$ states in the projectile $h$.
The $B$ results are frame independent, the $J/\psi$ and $\psi \psi$ averages
are in the laboratory frame.   The invariant masses are given in GeV.
Recall that $B_F$ denotes $B$ mesons produced
by fragmentation while $B_C$ denotes mesons produced by coalescence with
projectile valence quarks.  Note that the combined
$\pi^- N \rightarrow \psi \psi X$ data give $\langle x_\psi \rangle = 0.33$,
$\langle x_{\psi \psi} \rangle = 0.66$, and $\langle M_{\psi \psi} \rangle
= 7.4$ GeV while the average $\psi \psi$ mass from the $pN \rightarrow \psi
\psi X$ measurements is $\langle M_{\psi \psi} \rangle = 6.8$ GeV.}
\begin{tabular}{ccccccccccc}
& \multicolumn{2}{c}{$B_F$} & \multicolumn{2}{c}{$B_C$} &
\multicolumn{2}{c}{$B_F \overline B_F$} & \multicolumn{2}{c}{$B_F \overline
B_C$} & \multicolumn{2}{c}{$B_C \overline B_C$} \\
$h$, $p_{\rm beam}$ (GeV) &  $\langle x_B \rangle$ & $\langle x_\psi
\rangle$ & $\langle x_B \rangle$ & $\langle x_\psi \rangle$ &
$\langle x_{B \overline B} \rangle$ & $\langle x_{\psi \psi} \rangle$ &
$\langle x_{B \overline B} \rangle$ & $\langle x_{\psi \psi} \rangle$ &
$\langle x_{B \overline B} \rangle$ & $\langle x_{\psi \psi} \rangle$ \\
\tableline
$\pi^-$, 150 & 0.33 & 0.35 & 0.50 & 0.44 & 0.64 & 0.71 & 0.80 & 0.76 & 1.0 &
0.81 \\
$\pi^-$, 280 & 0.33 & 0.31 & 0.50 & 0.40 & 0.64 & 0.60 & 0.80 & 0.67 & 1.0 &
0.75 \\
$p$, 400 & 0.29 & 0.26 & 0.43 & 0.35 & 0.55 & 0.51 & 0.69 & 0.58 & & \\
$p$, 800 & 0.29 & 0.24 & 0.43 & 0.32 & 0.55 & 0.44 & 0.69 & 0.52 & & \\
\tableline
& & & & & $\langle M_{B \overline B} \rangle$ & $\langle M_{\psi \psi} \rangle$
&  $\langle M_{B \overline B} \rangle$ & $\langle M_{\psi \psi} \rangle$
&  $\langle M_{B \overline B} \rangle$ & $\langle M_{\psi \psi} \rangle$ \\
$\pi^-$ & & & & & 12.5 & 7.65 & 12.2 & 7.5 & 11.8 & 7.05 \\
$p$ & & & & & 12.5 & 7.65 & 12.4 & 7.6 & & \\
\end{tabular}
\end{table}

\widetext
\begin{table}
\caption{The average momentum fractions for $B$ production and decay
from an intrinsic $c \overline c
b \overline b$ state. The $\psi \psi$ pairs are produced through
$c \overline c
\rightarrow J/\psi$ coalescence and $B \rightarrow J/\psi X$ decay.
The $B$ averages are
frame independent, the $J/\psi$ and $\psi \psi$ averages are in the
laboratory frame.  Recall that $B_F$ denotes $B$ mesons produced
by fragmentation while $B_C$ denotes mesons produced by coalescence with
projectile valence quarks.  Note that the combined
$\pi^- N \rightarrow \psi \psi X$ data give $\langle x_\psi \rangle = 0.33$,
$\langle x_{\psi \psi} \rangle = 0.66$.}
\begin{tabular}{ccccccc}
& \multicolumn{2}{c}{$B_F$} & \multicolumn{2}{c}{$B_C$} &
$\psi B_F$ & $\psi B_C$ \\
$h$, $p_{\rm beam}$ (GeV) &  $\langle x_B \rangle$ & $\langle x_\psi
\rangle$ & $\langle x_B \rangle$ & $\langle x_\psi \rangle$ &
$\langle x_{\psi \psi} \rangle$ & $\langle x_{\psi \psi} \rangle$ \\
\tableline
$\pi^-$, 150 & 0.23 & 0.30 & 0.35 & 0.36 & 0.69 & 0.74 \\
$\pi^-$, 280 & 0.23 & 0.25 & 0.35 & 0.32 & 0.62 & 0.67 \\
$p$, 400 & 0.21 & 0.22 & 0.32 & 0.28 & 0.54 & 0.60 \\
$p$, 800 & 0.21 & 0.19 & 0.32 & 0.25 & 0.49 & 0.56 \\
\end{tabular}
\end{table}

\mediumtext
\begin{table}
\caption{A summary of $\sigma_{\psi \psi}$ for all the studied
processes.  When only one value is given for more than one
energy, the results are nearly energy independent.}
\begin{tabular}{ccccc}
Process & \multicolumn{2}{c}{$\sigma_{\psi \psi}(\pi^- N)$ (pb)} &
\multicolumn{2}{c}{$\sigma_{\psi \psi}(p N)$ (pb)} \\
& 150 GeV & 280 GeV & 400 GeV & 800 GeV \\
\tableline
NA3 & $18\pm8$ & $30\pm10$ & $27\pm10$ & \\
${\cal O}(\alpha_s^4)$ & 7.4 & 16.5 & 7.2 & 26.0 \\
${\cal O}(\alpha_s^2)$ $b \overline b$ & 0.028 & 0.24 & 0.03 & 0.16 \\
$2^{++} c \overline c c \overline c$ & 1.7 & 10.0 & 12.8 & 55.2 \\
$|n_V b \overline b \rangle$ & & 0.36 & & 0.51 \\
$|n_V c \overline c c \overline c \rangle$ & & $\equiv 20$ & & 6.1 \\
$|n_V c \overline c b \overline b \rangle$ & & 0.044 & & 0.020 \\
$|n c \overline c \rangle$ & 3.3 & 4.3 & 1.9 & 2.8 \\
\end{tabular}
\end{table}


\newpage
\begin{references}

\bibitem{Badpi}
J. Badier {\it et al.}, Phys. Lett. {\bf 114B} (1982) 457.

\bibitem{Badp}
J. Badier {\it et al.}, Phys. Lett. {\bf 158B} (1985) 85.

\bibitem{ES}
R.E. Ecclestone and D.M. Scott, Phys. Lett. {\bf 120B} (1983) 237.

\bibitem{HM}
B. Humpert and P. Mery, Phys. Lett. {\bf 124B} (1983) 265.

\bibitem{Russ}
V.G. Kartvelishvili and Sh.M. \'{E}sakiya, Sov. J. Nucl. Phys. {\bf 38}(3)
(1983) 430 [Yad. Fiz. {\bf 38} (1983) 722].

\bibitem{BHK}
V. Barger, F. Halzen, and W.Y. Keung, Phys. Lett. {\bf 119B} (1982) 453.

\bibitem{LL}
B.-A. Li and K.-F. Liu, Phys. Rev. {\bf D29} (1984) 426.

\bibitem{intc} S.J. Brodsky, P. Hoyer, C. Peterson, and N. Sakai,
Phys. Lett. {\bf B93} (1980) 451; S.J. Brodsky, C. Peterson and N.
Sakai, Phys. Rev. {\bf D23} (1981) 2745.

\bibitem{VB2}
R. Vogt and S.J. Brodsky, LBL-36754, SLAC-PUB-95-6753,
submitted to Phys. Lett. {\bf B}.

\bibitem{GRV}
M. Gluck, E. Reya, and A. Vogt, Z. Phys. {\bf C53} (1992) 127;

\bibitem{GRVpi}
M. Gluck, E. Reya, and A. Vogt, Z. Phys. {\bf C53} (1992)
651.

\bibitem{DO11}
J.F. Owens, Phys. Lett. {\bf B266} (1991) 126.

\bibitem{OW1}
J.F. Owens, Phys. Rev. {\bf D30} (1984) 943.

\bibitem{NDE}
P. Nason, S. Dawson, and R.K. Ellis, Nucl. Phys. {\bf B327} (1989) 49.

\bibitem{BaierR}
R. Baier and R. R\"{u}ckl, Z. Phys. {\bf C19} (1983) 251.

\bibitem{Schuler}
G. Schuler, CERN preprint CERN-TH.7170/94, submitted to Phys. Rep.

\bibitem{Ellis}
R.K. Ellis, in {\it Physics at the 100 GeV Scale},
Proceedings of the 17$^{\rm th}$ SLAC Summer Institute, Stanford, California,
1989, edited by E.C. Brennan (SLAC Report No. 361, Stanford, 1990).

\bibitem{PDG}
M. Aguilar-Benitez, Particle Data Group, Phys. Rev. {\bf D50} (1994) 1173.

\bibitem{Pete}
C. Peterson, D. Schlatter, I. Schmitt, and P. Zerwas, Phys. Rev. {\bf D27}
(1983) 105.

\bibitem{VBH2}
R. Vogt, S.J. Brodsky, and P. Hoyer, Nucl. Phys. {\bf B383} (1992) 643.

\bibitem{Chrin}
J. Chrin, in {\it Proc. Int. Symp. on Production and Decay of Heavy Flavors},
Stanford, California, 1987, edited by E. Bloom and A. Fridman, p. 131.

\bibitem{CLEO}
R. Balest {\it et al.}, CLEO Coll., CLNS-94-1315.

\bibitem{BHMT} S.J.  Brodsky, P. Hoyer, A.H. Mueller, and W.-K. Tang,
Nucl.  Phys. {\bf B369} (1992) 519.

\bibitem{BH}
P. Hoyer and S.J. Brodsky,  Nashville Part. Prod. 1990, p. 238.

\bibitem{EMCic}
J.J. Aubert {\it et al.}, Phys. Lett. {\bf 110B} (1982) 73;
E. Hoffmann and R. Moore, Z. Phys. {\bf C20} (1983) 71.

\bibitem{769}
M. Adamovich {\it et al.}, Phys. Lett. {\bf B305} (1993) 402;
G.A. Alves {\it et al.}, Phys. Rev. Lett. {\bf 72} (1994) 812.

\bibitem{VB}
R. Vogt and S.J. Brodsky, LBL-35380, SLAC-PUB-6468,
Nucl. Phys. {\bf B}, in press.

\bibitem{Bad}
J. Badier {\it et al.}, Z. Phys. {\bf C20} (1983) 101.

\bibitem{VBH1}
R. Vogt, S.J. Brodsky, and P. Hoyer, Nucl. Phys. {\bf B360} (1991) 67.

\end{references}
\end{document}